\documentclass[aps,prl,twocolumn,showpacs,groupedaddress,
 amsmath,amssymb,
]{revtex4-1}

\usepackage{graphicx} 
\usepackage{gensymb}
\usepackage{amsmath}
\usepackage{bm}

\usepackage{xcolor}
\usepackage{hyperref}
\hypersetup{
    urlbordercolor=black, 
  colorlinks   = true, 
  urlcolor     = blue, 
  linkcolor    = blue, 
  citecolor   = blue 
}

\begin{document}
\title{Coupling of non-crossing wave modes in a two-dimensional plasma crystal}

\author{J. K. Meyer}
\email{John.Meyer@dlr.de}
\author{I. Laut}
\author{S. K. Zhdanov}
\author{V. Nosenko}
\email{V.Nosenko@dlr.de}
\author{H. M. Thomas}
\affiliation{Institut f\"ur Materialphysik im Weltraum, Deutsches Zentrum f\"ur Luft- und Raumfahrt (DLR), 82234 We{\ss}ling, Germany}

\date{\today}
\begin{abstract}
We report an experimental observation of coupling of the transverse vertical and longitudinal in-plane dust-lattice wave modes in a two-dimensional complex plasma crystal in the absence of mode crossing. A new large diameter rf plasma chamber was used to suspend the plasma crystal. The observations are confirmed with molecular-dynamics simulations. The coupling manifests itself in traces of the transverse vertical mode appearing in the measured longitudinal spectra and vice versa.
We calculate the expected ratio of the trace to the principal mode with a theoretical analysis of the modes in a crystal with finite temperature and find good agreement with the experiment and simulations.
\end{abstract}
 \maketitle

\emph{Introduction.}
A complex, or dusty, plasma is a weakly ionized gas in which micrometer sized particles are immersed \cite{chu1994, *thomas1994, *hayashi1994, *thomas1996}. Due to absorption of electrons and ions the particles acquire a large charge. The high charge leads to strong coupling between the particles.  In ground-based experiments the particles become suspended in a two-dimensional (2D) layer in the sheath of the lower electrode, where the downward gravitational force is balanced by the upward electric force. The ability to directly image the particle motion via laser illumination and high speed videography allows the complete measurement of the state of the entire particle ensemble. Combining these factors, complex plasmas are convenient systems for experimental study of strongly coupled phenomena at the kinetic level \cite{ivlev2012, morfill2009, fortov2005}.

In a 2D complex plasma crystal, two in-plane dust-lattice (DL) wave modes are permitted, namely the longitudinal (L) and transverse horizontal (TH) modes with acoustic dispersion. Due to finite vertical confinement, a transverse vertical (TV) mode is also permitted \cite{vladimirov1997,ivlev2000,qiao2003,samsonov2005}. This vertical mode has negative optical dispersion and depends on the \emph{plasma wake} below the particles \cite{couedel2009}. The plasma wake is formed because in the sheath, the ions are accelerated toward the electrode by the electric field. This creates a flow of ions that interacts with the negatively charged particles. The ions are focused below the particles and create areas with higher density of positive charge. The plasma wake can be modeled as a point charge below the particle. It interacts with neighboring particles and is tied positionally relative to its seed particle \cite{melzer1996,lampe2000}. 

When studying generic phenomena like transport phenomena \cite{nosenko2008heat}, phase transitions \cite{killer2016} and linear \cite{misawa2001} and nonlinear waves \cite{tsai2016} in complex plasmas, care has to be taken that the plasma-specific processes do not play a role. One prime example of such a specific process is the so-called mode-coupling instability (MCI), where the L and TV modes intersect and at the intersection the unstable hybrid mode is formed which grows exponentially with time until the crystal melts \cite{zhdanov2009,couedel2010, laut2016anisotropic,roeker2012}. During the MCI, a \emph{mixed polarization} can be observed in experiments, where traces of the L mode can be measured in the transverse vertical spectra and vice versa \cite{liu2010, couedel2016}. It was believed that the mixed polarization could only be observed in the presence of mode crossing \cite{ivlev2012}.

In this Letter, we report on the observation of mixed polarization of L and TV modes in a 2D complex plasma crystal in the absence of mode crossing. Mutual coupling of wave modes with longitudinal and transverse polarization (and the emergence of mixed polarization) is important in other fields as well. One prominent example is surface plasmon polariton (SPP), which is a surface wave traveling along a metal-dielectric interface, where the surface plasmon in the metal and polariton in the dielectric are intrinsically coupled. SPPs are crucial in the physics of evanescent waves in a dense plasma \cite{fourkal2006}, surface waves guided by the interface of a metal and a dielectric material \cite{erten2017}, and in 2D nanoparticle arrays in photonics \cite{backes2008}.

\begin{figure} 
\centering
\includegraphics[width=\columnwidth]{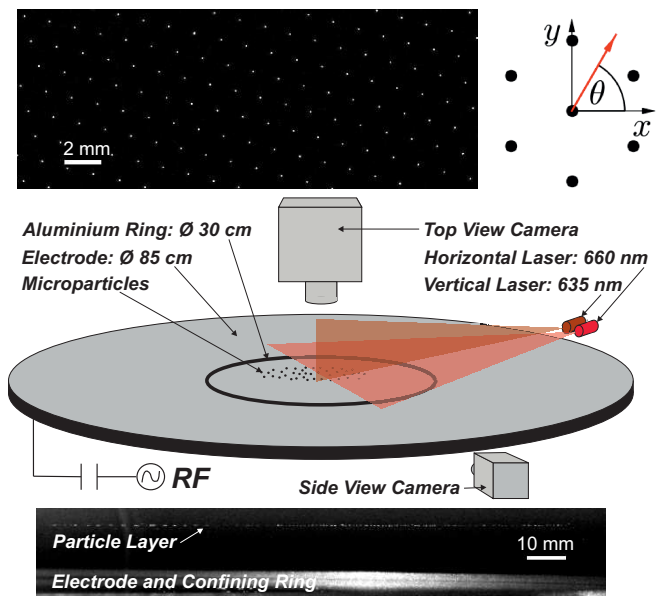}
\caption {\label {fig_exp_setup}
Layout of the setup used in the experiment with a cropped top view image (top left) from the experiment showing part of a crystal and a side view (bottom). The electrode is $85$ cm in diameter, and there is an aluminium ring with inner diameter $30$ cm for radial confinement. The microparticles are illuminated horizontally by a 660 nm wavelength laser sheet and vertically (through the center of the cloud) by a 635 nm wavelength laser sheet. The side view composite of the monolayer shows the highest intensity for each pixel from a stack of 10 frames. Due to a slight rotation of the crystal \cite{konopka2000}, this allows a larger number of particles to be shown. The left edge of the side view is at approximately the center of the layer and extends past the right edge of the layer. Just below the microparticles is a faint reflection of them in the window glass. For this paper, $\theta$ is defined as the angle in the plane measured from the midpoint of the nearest neighbors (top right).}
\end{figure} 

\emph{Experiment.}
For these experiments, a new large diameter plasma chamber was used (to be described fully in another paper). The powered electrode is $85$ cm in diameter and sustains a capacitively coupled rf glow discharge at $13.56$ MHz. The plasma is in argon at between $0.10$ and $1.00$ Pa. The rf power was set between $25$ and $200$ W at the power supply. Melamine formaldehyde spheres of diameter $9.19 \pm 0.09$ $\mu$m  were injected into the plasma and settled into a 2D suspension in the sheath above the electrode. They formed a crystalline structure of about $27$ cm diameter (due to a $30$ cm inner diameter ring placed on the bottom electrode that provided radial confinement). The particles were illuminated by a horizontal laser sheet with a wavelength of 660 nm and imaged from above through a matched interference filter by a $4$ megapixel camera at a speed of $60$ frames per second. A vertical cross section of the cloud was illuminated vertically by a 635 nm laser sheet and imaged through a matched interference filter. Using a standard particle tracking technique that finds the center of intensity of the particle image, the $x$ and $y$ position of the particles were measured in each frame with sub-pixel resolution. The velocities are then calculated by measuring the displacement of each particle between frames. The magnitude of the vertical displacements and velocities were estimated by measuring the brightness of each particle in each frame \cite{couedel2009}. The $x$ and $y$ axes are defined as depicted in Fig.~\ref{fig_exp_setup}, i.e., the $x$ axis falls on the line of the midpoint between two adjacent nearest neighbors.

\begin{figure} 
\centering
\includegraphics[width=\columnwidth]{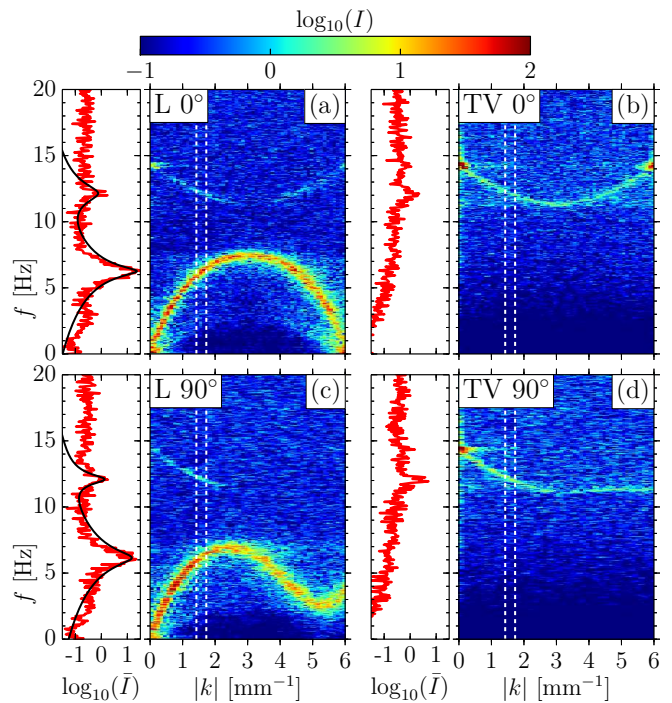}
\caption { \label{fig_exp_fluct_spect}
Experimental fluctuation spectra for longitudinal (L) and transverse vertical (TV) waves with wave vectors along $\theta = 0^\circ$ and $\theta=90^\circ$. To the left of each panel, the average intensity $\bar I$ in the range $1.4~\mathrm{mm}^{-1} < |k| < 1.8~\mathrm{mm}^{-1}$ is shown (see text for explanation of the fitted curves). Intensities are shown in arbitrary units and with logarithmic scale.}
\end{figure} 

From the particle positions and velocities, the particle current fluctuation spectra for each wave mode can be obtained for the experiment \cite{donko2008}. The spectra are calculated in the direction of the $x$ axis ($\theta = 0^\circ$) and in the direction of the $y$ axis ($\theta = 90^\circ$). The intensity of these spectra shows the wave energy distribution as a function of frequency $f$ and wave number $k$ for each mode.  In the L mode spectra shown in Figs.~\ref{fig_exp_fluct_spect}(a) and \ref{fig_exp_fluct_spect}(c), a trace of the TV mode can be clearly seen at $f = 11-15~\mathrm{Hz}$, despite the fact that the modes do not cross. The gap between the modes is $\simeq 4~\mathrm{Hz}$. In the experiment, the argon pressure was 0.15 Pa and the rf power was 150 W.
To quantify the strength of the mixed polarization, the intensity of the longitudinal spectra is averaged in the range $1.4~\mathrm{mm}^{-1} < |k| < 1.8~\mathrm{mm}^{-1}$ and fitted to two Lorentzians $l_{1,2} = A_{1,2} (\gamma_{1,2}/\pi)/( (f - f_{1,2})^2 + \gamma_{1,2}^2)$. The fitted curves shown in Figs.~\ref{fig_exp_fluct_spect}(a) and \ref{fig_exp_fluct_spect}(c) approximate the intensity near the modes well, but clearly deviate from the measured intensity where the noise dominates the spectrum. The \emph{mixed polarization ratio}, $R_m = A_1/A_2$, of the smaller to the larger amplitude is $R_m = 0.07 \pm 0.04$ for $\theta = 0^\circ$ and $R_m = 0.06 \pm 0.03$ for $\theta = 90^\circ$.

In the spectra for the TV mode shown in  Figs.\ \ref{fig_exp_fluct_spect}(b) and \ref{fig_exp_fluct_spect}(d), there is no evidence of a trace of the L mode. However, this is likely due to the low signal-to-noise ratio for this data which is on the order of the mixed polarization ratio in the longitudinal spectra.

Next, second order polynomial fits were made to the fluctuation spectra at low wave numbers and the slopes of the L and TH modes were calculated at zero wave number. This gives us the sound speed for each wave mode, and we calculate the charge $Q = -35000 \pm 2000~e$ and screening parameter $\kappa = a/\lambda = 1.22 \pm 0.18$ using the method of Ref.~\cite{nunomura2002dispersion}. $a = 1.18 \pm 0.11~\mathrm{mm}$ is the interparticle distance and $\lambda$ the screening length of the Yukawa interaction. The particle kinetic temperature is $2.9 \pm 0.6$ eV and the effective coupling parameter $\Gamma^{*}$, defined in Ref.~\cite{hartmann2005}, is $800 \pm 140$.

\emph{Simulation.}
A trivial reason for the observation of the mixed polarization in experiments could be the geometric effect: When observing the crystal not perfectly from above but at an angle, the vertical particle displacement is also projected onto the measured $xy$ displacement. A bending of the crystal could be a different reason for the mixed polarization; from the side view (see Fig.~\ref{fig_exp_setup}) it can be observed that the vertical deflection of the layer is less than 0.5 mm over the 150 mm observed. As the geometric effect can be completely excluded in simulations, and the crystal bending assured to be negligible, we performed molecular dynamics simulations with inputs based on the measured experimental values.
The equation of motion for particle $i$ reads:
\begin{equation}\label{eq_of_motion}
M\ddot{\mathbf{r}}_i + M \nu \dot{\mathbf{r}}_i = \sum_{j \neq i} \mathbf{F}_{ji} - \bm{\nabla}V_i + \mathbf{L}_i,
\end{equation}
where $\mathbf{r}_i$ is the three-dimensional particle position, $M$ the mass and $\nu $ the damping rate. The forces acting on the particle are the mutual particle interactions $\mathbf{F}_{ji}$, the force derived from the external potential $V_i$, and a Langevin heat bath $\mathbf{L}_i$.

To include the ion wake in the mutual particle interaction, a positive pointlike charge $q$ is placed a fixed vertical distance $\delta$ below each particle, while the particle itself is modeled as a negative pointlike charge $Q < 0$. The force exerted by particle $j$ (and its wake) on particle $i$ is thus modeled as
\begin{equation}
\label{eq_interparticel_forces}
\mathbf{F}_{ji} =                                     
 Q^2 f( r_{ji} ) \frac{\mathbf{r}_{ji}}{r_{ji}}
 +qQ f( r_{ w_{ji} } ) \frac{ \mathbf{r}_{ w_{ji} } }{ r_{ w_{ji} } }, \\ 
\end{equation}
where $f(r) = \exp ( -r/\lambda) ( 1 + r/\lambda) / r^2$, $\lambda$ the screening length,
$\mathbf{r}_{ji} = \mathbf{r}_{i} - \mathbf{r}_{j}$ and
$\mathbf{r}_{w_{ji}} = \mathbf{r}_{i} - (\mathbf{r}_{j} - \delta \mathbf{e}_z)$. Here and in the following, $r$ denotes the magnitude of vector $\mathbf{r}$, and $\mathbf{e}_{x,y,z}$ are the unit vectors of the coordinate system. The magnitude and distance of the wake charge can be described by the dimensionless parameters $\widetilde q = q / |Q|$ and $\widetilde \delta  = \delta /\lambda$.

The external potential in Eq.~\ref{eq_of_motion} reads $V_i = 0.5 M \left( \Omega_h^2 \rho_i^{10}/R^8 + \Omega_z^2 z_i^2 \right)$, where $\rho_i = \sqrt{x_i^2 + y_i^2}$ is the horizontal position of particle $i$, $\Omega_h$ and $\Omega_z$ the horizontal and vertical confinement frequencies and  $R$ the approximate horizontal radius of the crystal. The potential mimics the strong vertical confinement due to the counterdirected electric and gravitational forces and the weaker horizontal confinement. The tenth-order dependence of $V_i$ on $\rho$ leads to a potential that is very flat in the horizontal direction for $\rho < R$ such that a very homogeneous crystal can be simulated \cite{durniak2010}.

The Langevin force $\mathbf{L}_i(t)$ is defined by $\langle \mathbf{L}_i(t) \rangle = 0$ and $\langle \mathbf{L}_i(t + \tau) \mathbf{L}_j(t) \rangle = 2 \nu M T \delta_{ij} \delta(\tau)$, where $T$ is the temperature of the heat bath, $\delta(t)$ the delta function and $\delta_{ij}$ the Kronecker delta.

In the simulations, the $N = 10000$ particles each had a mass of $M = 0.61 \times 10^{-12}$~kg and charge $Q=-49295~e$. The screening length was $\lambda = 2.4~\mathrm{mm}$ and $\nu = 1~\mathrm{s}^{-1}$. The wake parameters were $\widetilde q = 0.75$ and $\widetilde \delta = 0.2$, yielding an effective particle charge of $Q_\mathrm{eff} = Q\sqrt{1-\widetilde q} \simeq -24648~e$. The parameters of the confinement were $\Omega_h = 2\pi \times 0.12~\text{s}^{-1}$, $\Omega_z = 2\pi \times 14~\text{s}^{-1}$, and $R = 63$~mm, yielding an interparticle distance of $a = 1.217\pm0.006~\mathrm{mm}$ in the suspension center. The temperature of the heat bath was $T = 300$ K.

\begin{figure} 
\centering
\includegraphics[width=\columnwidth]{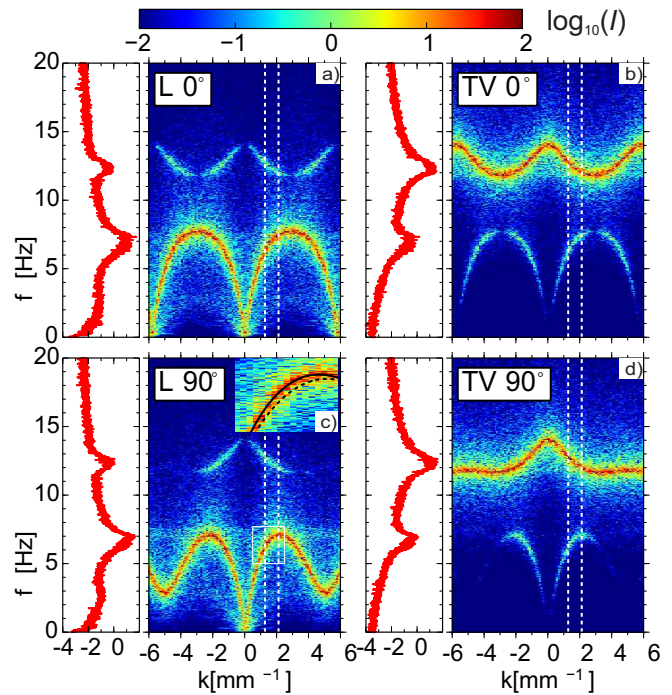}
\caption { \label {fig_sim_fluct_spect}
Same as Fig.~\ref{fig_exp_fluct_spect} but for the simulated crystal. The mixed polarization ratio is $R_m = 0.08 \pm 0.02$ (a), $R_m = 0.09 \pm 0.02$ (b), $R_m = 0.08 \pm 0.02$ (c), $R_m = 0.09 \pm 0.02$ (d). The inset shows a zoom in the region within the white frame, together with the theoretical coupled (solid line) and uncoupled (dashed line) modes \cite{couedel2011}.}
\end{figure} 

As shown in Fig.~\ref{fig_sim_fluct_spect}, the mixed polarization is clearly seen in the simulations. The mixed polarization ratio is $R_m = 0.08\pm0.02$ for the L mode. Within the accuracy, it is identical in the two directions $\theta = 0^\circ$ and $\theta = 90^\circ$. As in the simulations the vertical particle positions can be known to arbitrary accuracy, the mixed polarization is also observed in the TV mode with a ratio of $R_m = 0.09\pm 0.02$. The mixed polarization is most pronounced for intermediate values of the wave number $|k|$, while it vanishes at $k=0$ and at the boundary of the first Brillouin zone.

In the inset in Fig.~\ref{fig_sim_fluct_spect}(c), a part of the L mode spectrum is shown together with the theoretical modes calculated as in Ref.~\cite{couedel2011}. Even in the absence of the mode-coupling instability, the coupled wave modes give a better agreement to the observed spectra than the uncoupled modes.

\emph{Theory.}
The equation of motion (Eq. \ref{eq_of_motion}) can also be analyzed theoretically. For an infinite horizontal crystal, only the vertical confinement has to be considered, such that the potential now reads  $V_i = 0.5 M \Omega_z^2 z_i^2$.

In the absence of noise ($\mathbf{L}_i = 0$), the crystal is a perfect hexagonal lattice where the equilibrium particle positions can be written as $\mathbf{r}_i^{(0)} \equiv \{\frac{\sqrt{3}}{2}ma,(n+\frac12 m)a,\Delta\}$.  The wake-mediated downward shift of the lattice plane reads
\begin{equation}\label{eq4}
\Delta=\frac{qQ\delta}{M\lambda\Omega_z^2}\sum_{m',n'\neq 0}\frac{s+\lambda}{s^3}\exp(-\frac{s}{\lambda})<0,
\end{equation}
where $s^2=a^2\left(m'^2+n'^2+m'n'\right)+\delta^2$. $\Delta$ is a convenient measure of the wake influence in simulations.

\begin{figure} 
\centering
\includegraphics[width=\columnwidth]{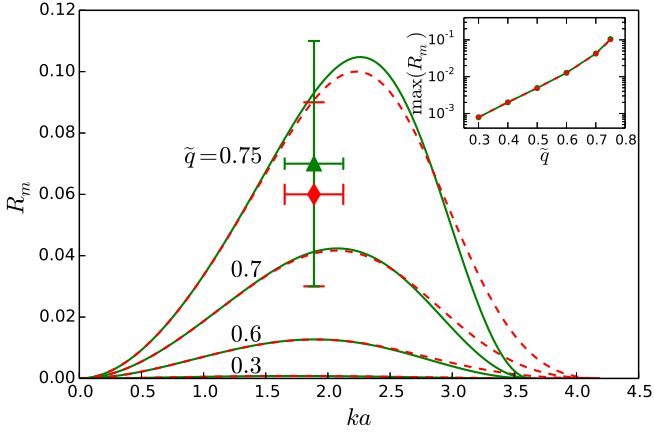}
\caption {\label {fig_theoretical_ratio} Mixed polarization ratio $R_m$ for $\theta=0\degree$ (solid line) and $\theta=90\degree$ (dashed line) with respect to the wave number $k$ (normalized by $a^{-1}$) for different values of $\widetilde q$. The triangle ($\theta = 0\degree$) and the diamond ($\theta = 90\degree$) show the measured $R_m$ from the experiment for the range of $ka$ indicated by the horizontal error bars. The inset shows the maximum of $R_m$ as a function of $\widetilde q$.}
\end{figure} 

A perturbed, linearized system, now at $\mathbf{L}_i \neq 0$, sustains both L and TH sound waves with in-plane polarization, as well as optical TV waves with vertical $z$-polarization \cite{zhdanov2009}. For example, the complex amplitudes $\widetilde{x},\widetilde{z}$ or $\widetilde{y},\widetilde{z}$ of the coupled L and TV waves proportional to $\exp[ik_xx^{(0)}+ik_yy^{(0)}]$ and propagating along the main crystallographic directions (either along $x$ ($\theta=0$) or $y$ ($\theta=\pi/2$); see inset in Fig.~\ref{fig_exp_setup}) could be described as
\begin{eqnarray}\label{eq5}
\ddot{\widetilde{\xi}}+\nu\dot{\widetilde{\xi}}=-\Omega_L^2\widetilde{\xi}+i\Omega_W^2\widetilde{z}+\widetilde{a}_\xi,\nonumber\\
\ddot{\widetilde{z}}+\nu\dot{\widetilde{z}}=-\Omega_V^2\widetilde{z}+i\Omega_W^2\widetilde{\xi}+\widetilde{a}_z,
\end{eqnarray}
where $\xi=x(y)$ when $\theta=0(\pi/2)$, and $\mathbf{\widetilde{a}}=\mathbf{L}_i/M$ are the time-dependent random accelerations.

The longitudinal (L) and vertical (V) eigenfrequencies, and the wake coupling (W) frequency are given by the following sums running over the integers $(m, n)$ with excluded $(0, 0)$:
\begin{equation}\label{eq14}
\begin{aligned}
\Omega_L^2 = &~ 2 \Omega_0^2 \sum_{m,n} \left[ \frac{\xi^2}{r_w^2} \left( \Xi_r - \Xi_w +\frac{\delta^2}{r_w^2} \Xi_w \right) - \Lambda_r + \Lambda_w \right] \\
             &~ \times \sin^2(k\xi/2), \\ 
\Omega_V^2 = &~ \Omega_z^2-2\Omega_{0}^2 \sum_{m,n}\left[\Lambda_r - \Lambda_w + \frac{\delta^2}{r_w^2}\Xi_w \right] \sin^2\left(k \xi/2\right), \\
\Omega_W^2 = &~\Omega_0^2 \delta\sum_{m,n} \Xi_w\frac{\xi}{r_w^2}\sin\left(k\xi\right), \\
\end{aligned}
\end{equation}
where $\Omega_{0}^2=Q^2/(M\lambda^3)$ is the dust lattice frequency, $\xi=\sqrt{3}am/2$ for $\theta=0$, and $\xi= a(n+m/2)$ for $\theta=\pi/2$. In Eq.~\ref{eq14}, $\Xi_r = \Xi(r/\lambda)$ and $\Xi_w = \widetilde{q} \Xi(r_w/\lambda)$, and likewise for $\Lambda_{r,w}$, where
$\Xi(x)=\frac{3+3x+x^2}{x^3}e^{-x}$ and $\Lambda(x)=\frac{1+x}{x^3}e^{-x}$.
The distances $r$ and $r_w$ are expressed in terms of $m$, $n$ as $r=a\sqrt{m^2+n^2+mn}$ and $r_w=\sqrt{r^2+\delta^2}$.

The frequencies $\Omega_{L,V,W}$ only depend on $k$ and $\theta$ but not on time. Therefore, the system (\ref{eq5}) of linear equations with constant coefficients is readily solved by performing corresponding averaging to find the spectral intensities $I_{L,V}(k,\omega)\propto\langle|\widetilde{v}_{\xi,z}|^2\rangle$ of coupled natural wave modes. The result, omitting details of these simple but rather tedious computations, is:
\begin{equation}\label{eq7}
I_{\{L,V\}}=I_0\frac{\omega^2\left[\left|\Omega^2-\Omega_{\{V,L\}}^2\right|^2+\Omega_W^4\right]}
{\left|\left[\Omega^2-\Omega_V^2\right]\left[\Omega^2-\Omega_L^2\right]+\Omega_W^4\right|^2},
\end{equation}
where (for brevity) $\Omega^2\equiv\omega(\omega+i\nu)$ and $I_0$ is a prefactor independent of $k$ and $\omega$. As the damping rate is rather weak (as in the present experiment), both $L$- and $V$-mode intensity distributions exhibit two narrow sharp maxima located at $\omega=\omega_{1,2}(k)$ (the normal modes) that could be introduced as:
\begin{equation}\label{eq8}
\omega_{1,2}^2=\frac12\left[\Omega_V^2+\Omega_L^2\pm\left(\Omega_V^2-\Omega_L^2\right)\sqrt{1-p}\,\right],
\end{equation}
where $p=4\Omega_W^4/(\Omega_V^2-\Omega_L^2)^2$.
Since normally $\Omega_V>\Omega_L$, the upper branch $\omega_1$ is close to the vertical eigenfrequency, but slightly smaller, while the lower one is close to the eigenfrequency of the $L$-mode. The 'redundant' peak at the $V$-mode frequency appearing in the $L$-mode spectrum (and vice versa) is a direct consequence of the mode coupling. The mixed polarization ratio can then be predicted theoretically as a function of the $p$-factor as
\begin{equation}\label{10}
R_m = \frac{I_L|_{\omega=\omega_1}}{I_L|_{\omega=\omega_2}} = \frac{I_V|_{\omega=\omega_2}}{I_V|_{\omega=\omega_1}} = \frac{p}{\left(1+\sqrt{1-p}\right)^2}+\mathcal{O}(\nu^2).
\end{equation}
The zero wake condition $\widetilde q = 0$ corresponds to $p = 0$. The mixed polarization ratio $R_m$, shown in Fig.~\ref{fig_theoretical_ratio}, was calculated using the parameters of the simulations for $\theta = 0^\circ$ and $\theta = 90^\circ$ and for different values of the wake charge $\widetilde q$. It is obvious that the mixed polarization is a wake-mediated effect which can only be observed for relatively large values of $\widetilde q$.
The value of $R_m$ observed in the experiment is reproduced for $0.7 < \widetilde q < 0.75$.
In good agreement with the experiment and simulation, $R_m$ is maximal for intermediate values of $k$ while it vanishes for $k=0$ and at the boundary of the first Brillouin zone.
The inset of Fig.~\ref{fig_theoretical_ratio} shows that $R_m$ grows almost exponentially with $\widetilde q$.
$R_m$ is evidently symmetric with respect to the L and TV modes, in good agreement with the simulations.

\emph{Summary.}
We observed in experiments mixed polarization between the longitudinal in-plane and transverse out-of-plane wave modes in the absence of mode crossing. Simulations show that this mixed polarization can be attributed to the ion wake. The theoretical analysis of the coupled modes matches the simulation and the experiment.

The theory predicts that there will not be mixed polarization observed in the TH mode and, in these experiments and simulations this is the case. Introducing the geometric effect artificially by rotating the simulated data about the $x$ or $y$ axis yields a spurious mixed polarization in both the L and TH modes. The geometric effect can thus be excluded for the experiments, where care was taken to only analyze the central field of view.

Our analysis showed that the mixed polarization can only be expected for relatively strong wake charges, which could explain why it has not been observed before. For example, for the value $\widetilde q = 0.3$ used in the simulations of Ref.~\cite{couedel2011}, a mixed polarization ratio of $R_m \simeq 10^{-3}$ could be expected from Fig.~\ref{fig_theoretical_ratio}. The small damping rate due to the low gas pressure of the present experiment and the high crystal quality are further reasons why the mixed polarization can be observed in the new plasma chamber.

\begin{acknowledgments}
The authors would like to thank Mierk Schwabe for helpful discussions. JM acknowledges the support of DLR-DAAD Research Fellowships.
\end{acknowledgments}

\end{document}